\documentclass[onecolumn,superscriptaddress,aps]{revtex4}
\usepackage[utf8]{inputenc}
\usepackage[T1]{fontenc}
\usepackage{rotating} 
\usepackage{graphicx,bm,bbm}
\usepackage{amsmath,amssymb}

\begin{document}

\title{Implementation of Magic State Injection within Heavy-Hexagon Architectures}
\author{Hansol Kim}
\email{khshk18@hanyang.ac.kr}
\author{Wonjae Choi}
\email{marchenw@hanyang.ac.kr}
\author{Younghun Kwon}
\email{yyhkwon@hanyang.ac.kr}
\affiliation{Department of Applied Physics, Hanyang University(ERICA), Ansan 15588, Republic of Korea}
\begin{abstract} 
\indent The magic state injection process is a critical component of fault-tolerant quantum computing, and numerous studies have been conducted on this topic. Many existing studies have focused on square-lattice structures, where each qubit connects directly to four other qubits via two-qubit gates. However, hardware that does not follow a lattice structure, such as IBM's heavy-hexagon structure, is also under development. In these non-lattice structures, many quantum error correction (QEC) codes designed for lattice-based system cannot be directly applied. Adapting these codes often requires incorporating additional qubits, such as flag qubits. This alters the properties of the QEC code and introduces new variables into the magic state injection process. In this study, we implemented and compared the magic state injection process on a heavy-hexagon structure with flag qubits and a lattice structure without flag qubits. Additionally, we considered biased errors in superconducting hardware and investigated the impact of flag qubits under these conditions. Our analysis reveals that the inclusion of flag qubits introduces distinct characteristics into the magic state injection process, which are absent in systems without flag qubits. Based on these findings, we identify several critical considerations for performing magic state injection on heavy-hexagon systems incorporating flag qubits. Furthermore, we propose an optimized approach to maximize the efficacy of this process in such systems. 
\end{abstract} 

\maketitle
%
%

\section{Introduction}
\indent Various types of quantum computers are currently under development, enabling numerous quantum operations. However, for a quantum computer to perform effective calculations, the error rate of the two-qubit gate must be reduced below a certain level.\cite{Moses, Evered, IonQ, IBM} For example, implementing quantum algorithms such as Shor's algorithm\cite{Shor}, --expected to outperform classical computers--requires a gate error rate between approximately $10^{-7}$ to $10^{-12}$\cite{Campbell, Kivlichan}. Achieving such low error rates solely through hardware improvements is nearly impossible. Therefore, quantum error correction (QEC) is essential to construct logical qubits and perform logical operations necessary for executing long and complex algorithms. \cite{Deutsch, Grover, Bae, Park, Simon, Vazirani, Harrow} The approach of performing computations using logical qubits while ensuring that the system remains resistant to errors is known as fault-tolerant quantum computing (FTQC). For FTQC, a QEC code with a high threshold error rate, compatible with a quantum computer, is required.\cite{Daniel, Roffe} To perform quantum computing, logical operators implementing arbitrary unitary operations that transform prepared logical states into other desired logical states are required.\cite{Horsman, Andrew, Christophe} Determining physical operations for all unitary operators during computation is inefficient. Instead, arbitrary unitary gates are implemented by combining specific gates with well-known physical operations. \cite{Bravi} The set of gates enabling arbitrary unitary operations is called a universal gate set. It can be constructed using various combinations of gates but must include at least one non-Clifford gate. \cite{Eastin}Although logical Clifford gates can be implemented using transversal gates directly on data qubits, logical non-Clifford gates cannot be represented in this manner.\cite{Zeng} To implement logical non-Clifford gates, an alternative method utilizing magic states has been proposed. \cite{Bravi, Li, Singh, Lao} Magic states are specialized states that enable the indirect implementation of logical non-Clifford gates. A distinct magic state exists for each non-Clifford gate to be implemented. Suppose that a logical qubit is prepared in the magic state. By applying only Clifford gates between this magic state qubit and the target logical qubit, the same results can be achieved as directly applying the non-Clifford gate to the target logical qubit.

\begin{figure}[t]

\centerline{\includegraphics[width=0.5\linewidth]{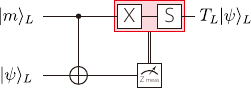}}

\caption{Logical quantum circuit for implementing a Logical T gate: If the state $\left| m \right\rangle_L = \left| 0 \right\rangle_L + e^{i\pi/4} \left| 1 \right\rangle_L$ is prepared, this circuit applies the Logical T gate to any arbitrary state using only logical Clifford gates. Since the magic state collapses due to measurement during the process, a new magic state is required for each logical T gate application.}

\label{fig:T gate}

\end{figure}

\indent For example, to implement the logical T gate, a representative non-Clifford gate, through magic state injection, the magic state \( |m\rangle_L = |0\rangle_L + e^{i\pi/4}|1\rangle_L \) must first be prepared.Applying the circuit shown in Figure \ref{fig:T gate} to a logical qubit prepared in this magic state and another logical qubit in the state \( |\psi\rangle_L \) (the target qubit for the logical T gate) produces the state corresponding to \( T_L|\psi\rangle_L \). Here, \( T_L \) denotes the logical T gate\cite{Lao}. To implement logical non-Clifford gates in this manner, the preparation of a logical magic state is essential. However, preparing a magic state from an available logical state, such as the logical 0 state ($|0\rangle_L$), using only logical Clifford gates is not feasible. Therefore, a magic state must be prepared in a physical qubit system, after which the state is encoded into a logical state to produce the desired logical magic state. This process of encoding this physical magic state into a logical magic state is called magic state injection. \\
\indent The magic state injection process varies depending on the QEC method used to implement the logical non-Clifford gate. The optimal QEC code for a given hardware depends on the characteristics of the hardware. Several quantum error correction codes, including the surface code\cite{Kitaev, Fowler, Bravyi2}, have been designed with a structure that allows direct connections to the four surrounding qubits. Google's quantum hardware is based on this structure and supports surface code implementations. \cite{Google, Google2}. However, not all hardware systems satisfy these conditions. IBM's hardware, which uses a heavy-hexagon structure, allows each qubit to connect directly to only 2-3 other qubits via two-qubit gates. \cite{Zhang}. This restriction makes it impossible to directly apply surface codes, which require connections up to four adjacent qubits for two-qubit gates. \\
\indent To address this limitation, additional qubits such as flag qubits must be used. \cite{Reichardt, Reichardt2, Almudever, Chamberland, Wu, Kim, Benito, Kim2}. Moreover, IBM's qubits typically exhibit $T_1$ times longer than $T_2$ times\cite{IBM}, indicating a Z bias, where Z errors are more likely than X errors when modeling errors using the Pauli error model. When such an error bias exists, other QEC codes, such as the XZZX code\cite{Ataides, Darmawan}, often outperform surface codes by exploiting this characteristic.Therefore, evaluating the hardware characteristics is essential to determine the most suitable QEC code. This evaluation ensures that the magic state injection process operates efficiently with the chosen code. \\
\indent In this study, we implemented the magic state injection process using surface and XZZX codes on the lattice and heavy-hexagon structures. We compared the logical error rates of the generated logical magic states while varying parameters, such as error bias, code distance, and state initialization. During this process, we observed several unique characteristics of the magic state injection process in heavy-hexagon structures with flag qubits. Additionally, we explored the most suitable magic state injection methods for heavy-hexagon hardware.
\section{Magic state Injection on Heavy-Hexagon Structure}
\subsection{QEC Code on Heavy-Hexagon Structure}

\begin{figure}[!t]
\centerline{\includegraphics[width=0.9\linewidth]{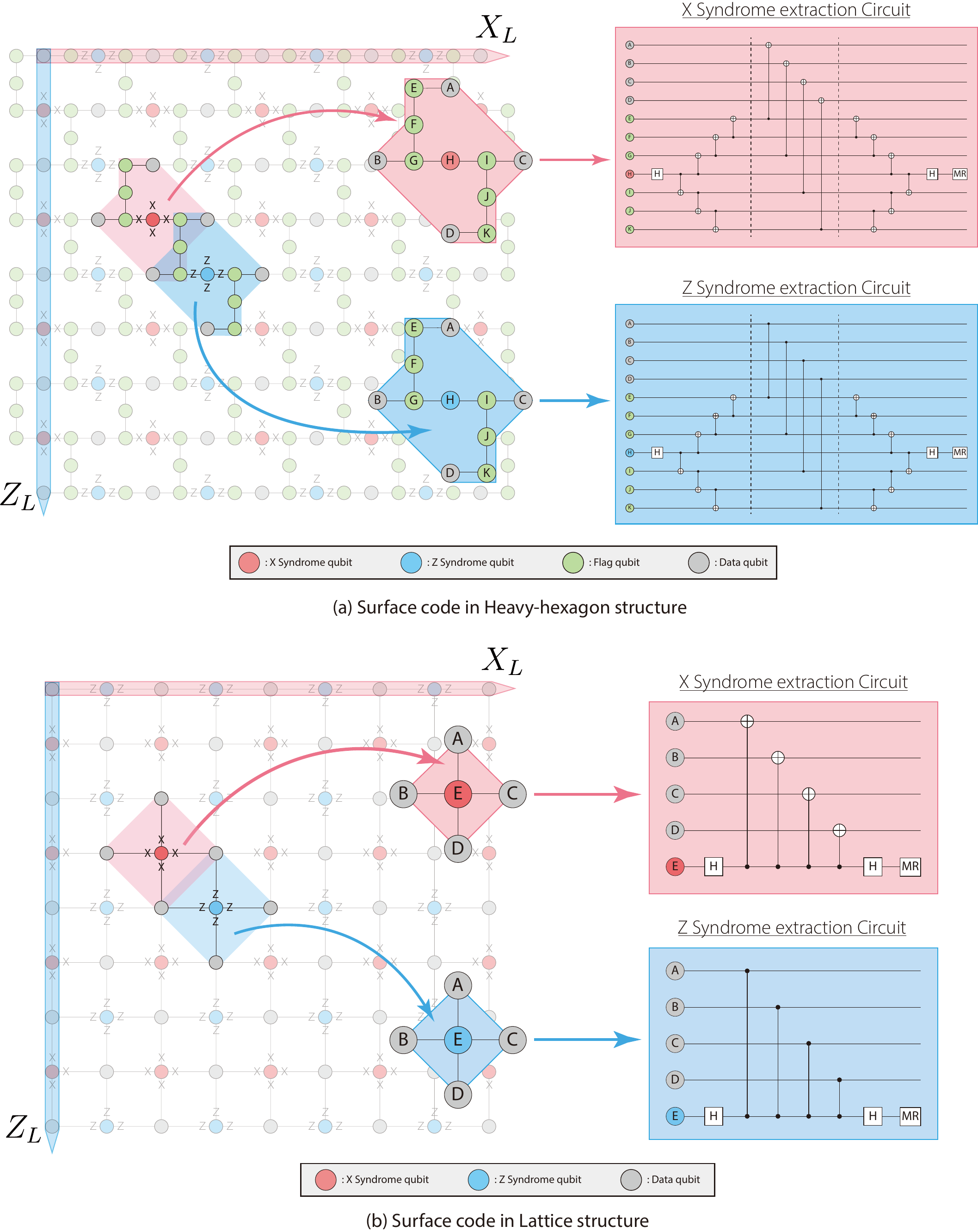}}
\caption{(a) Surface code on the heavy-hexagon structure. (b) Surface code on the lattice structure. Data qubits are shown in gray, X syndrome qubits in red, Z syndrome qubits in blue, and flag qubits in green. The red- and blue-shaded areas correspond to stabilizer operators. Flag qubits enable stabilizer measurements by indirectly connecting data qubits and syndrome qubits that cannot be directly connected through two-qubit gates in the heavy-hexagon structure. Except for the green flag qubits, the surface code on the heavy-hexagon structure is identical to that on the lattice structure.}
\label{fig:surface code on HH with flag}
\end{figure}

\indent The surface code typically requires each qubit to connect directly to four neighboring qubits. However, IBM's Heavy-hexagon structure limits connectivity to only two or three neighboring qubits. To address this limitation, we previously introduced additional qubits known as flag qubits. These qubits enable stabilizer measurements by indirectly connecting data qubits and syndrome qubits. This approach facilitates the implementation of error correction code on the heavy-hexagon structure, as illustrated in Figure \ref{fig:surface code on HH with flag}.

As shown in the figure, the surface code on the heavy-hexagon structure follows a layout similar to that of the surface code on the lattice structure, apart from the addition of flag qubits. The surface code used in the experiment includes two types of syndrome qubits. The X syndrome qubits (red) measure the stabilizers of the form $X_AX_BX_CX_D$, applying the Pauli X operator to the four neighboring data qubits. The Z syndrome qubits (blue) measure the stabilizers of the form $Z_AZ_BZ_CZ_D$, applying the Pauli Z operator to four neighboring data qubits. The logical X and Z operators are associated with chains of errors that extend from one boundary to the opposite boundary. Specifically, a chain of Z errors stretching from the top to the bottom boundary defines a logical Z operator. Similarly, a chain of X errors running from the left to the right boundary defines a logical X operator. Since the stabilizer operators remain unchanged, the logical operations for the surface code on the heavy-hexagon structure are consistent with those for the lattice structure. By introducing flag qubits, the traditional surface code requirement for four-qubit connectivity is effectively adapted for heavy-hexagon structure.

\begin{figure}[!t]

\centerline{\includegraphics[width=0.9\linewidth]{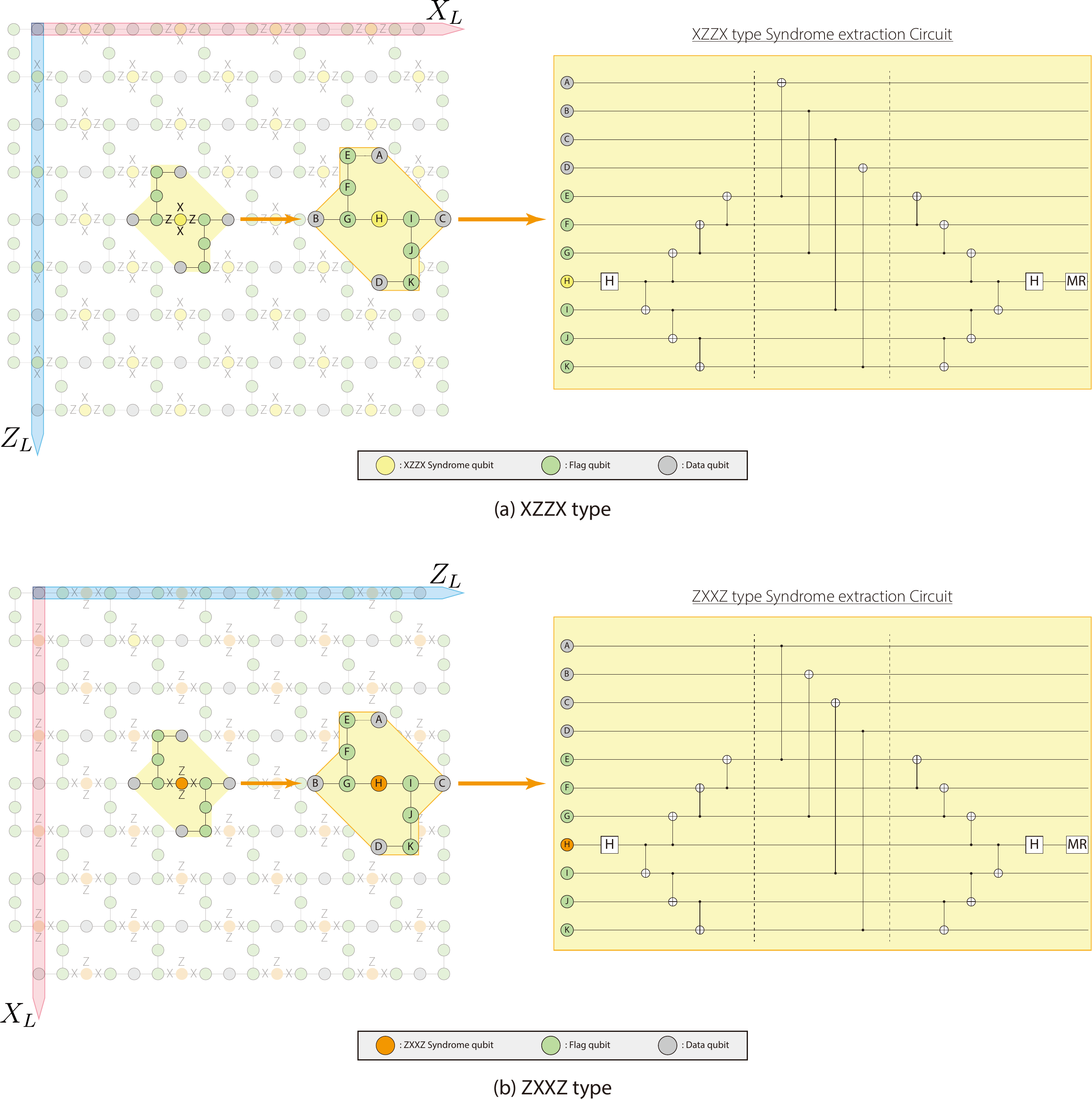}}

\caption{XZZX code on the heavy-hexagon structure. The XZZX code detects and corrects errors in logical state through a stabilizer composed of Pauli X and Pauli Z operators. There are two ways to arrange Pauli X and Pauli Z in the stabilizer. (a) XZZX type, which places Pauli X operators vertically and Pauli Z operators horizontally relative to the syndrome qubit. (b) ZXXZ type, which places Pauli Z operators vertically and Pauli X operators horizontally. In these two cases, the forms of the logical X and Z operators are reversed. In the figure, the logical X operator is highlighted in red, and the logical Z operator is highlighted in blue. In lattice structures, this difference does not typically affect the performance of the error correction code. However, in the heavy-hexagon structure using flag qubits, the number of flag qubits required to connect vertical and horizontal directions varies. This variance may influence the performance of error correction depending on the selected stabilizer configuration.}

\label{fig:XZZX code on HH with flag}
\end{figure}

\subsection{XZZX code}
\indent The XZZX code is an error correction code designed to detect both X and Z errors using a single type of stabilizer, which incorporates both Pauli X and Pauli Z operators. The stabilizer for the XZZX code takes the form \( X_AZ_BZ_CX_D \). In this configuration, Pauli X is applied to the data qubits positioned above and below the syndrome qubit. Meanwhile, Pauli Z is applied to the data qubits located on the left and right sides. (see Figure \ref{fig:XZZX code on HH with flag}). In this stabilizer structure, because X and Z errors propagate in perpendicular directions, the relative positions of the detected syndromes enable the identification of the error type. Since a single syndrome qubit detects both types of errors, prior information of an error bias can be utilized to infer whether the detected error is more likely an X or Z type error.  Leveraging this additional information increases the probability of determining the correct correction operator during decoding. Consequently, several studies have shown that the XZZX code performs better under biased error models. Similar to the previously described methods, the XZZX code can also be implemented on a heavy-hexagon structure by appropriately positioning the data, syndromes, and flag qubits. \\
\indent When applying the XZZX code on a heavy-hexagon structure, an additional consideration arises. In the previously described XZZX code, Pauli X is applied to the data qubits above and below the syndrome qubit, while Pauli Z is applied to the left and right data qubits. However, different stabilizers can also be constructed. A stabilizer of the form  $Z_AX_BX_CZ_D$, which applies Pauli Z above and below the syndrome qubits, and Pauli X to the left and right qubits. Both configurations detect and correct errors equivalently. In a regular lattice structure, these two stabilizers are symmetric, and the choice of configuration has no significant impact on performance. However, in the heavy-hexagon structure, the number of flag qubits needed to connect data  and syndrome qubits differs between the vertical and horizontal directions. This asymmetry leads to differences in performance depending on whether the stabilizer uses the XZZX or ZXXZ form. Therefore, we conducted experiments by dividing the XZZX code into two types: XZZX type and ZXXZ type. \\
\indent The logical operators of the XZZX code are defined similarly to those in the surface code. However, the stabilizer forms of the two types of XZZX code result in inverted logical operators. For example, the logical operators of the XZZX type are identical to those of the surface code. However, for the ZXXZ type, the logical operators are reversed. Specifically, in the ZXXZ type, a chain of X errors spanning from the top boundary data qubits to the bottom boundary data qubits represents a logical X error. Similarly, a chain of Z errors extending from the left boundary data qubits to the right boundary data qubits corresponds to a logical Z error. (see Figure \ref{fig:XZZX code on HH with flag}).

\subsection{Magic State Injection}

\indent The magic state injection is a process that maps a desired magic state implemented on physical qubits to the corresponding logical state of an error-correction code. This involves creating a state that satisfies two key conditions: (1) it must be an eigenstate of all the stabilizer operators in the error correction code, and (2) the expectation value of the logical Pauli operator for the logical state must match the expectation value of the physical Pauli operator for the physical qubit.  To achieve this, the eigenstate of the stabilizers is prepared through a projection process involving stabilizer measurements. Simultaneously, the physical state for  certain physical qubits must be encoded into the desired logical state. Since logical Pauli operators commute with all stabilizer operators, their expectation value can be made equal for both physical and logical states, provided the state is appropriately prepared. For instance, consider injecting an arbitrary magic state $|M\rangle = \alpha|0\rangle + \beta|1\rangle$ into a logical state. In this case, the $Z$ error in all the data qubits in the first row corresponds to a logical $Z$ error ($Z_L$). If the physical magic state is prepared on one of the qubits in the first row while initializing the other qubits in the same row to the $|0\rangle$, the expected value of $Z_L$ for this state matches that of $Z_L$ for the physical magic state. Because $Z_L$ commutes with all the stabilizers, this relationship remains valid even after performing stabilizer measurements. The initialized state can be expressed as follows, where $\left| \phi \right\rangle_{else}$ represents the state of the data qubits not involved in $Z_L$, and $d$ is the distance of the logical state being encoded:

\begin{align}
\left| \psi \right\rangle = \alpha \left| 0 \right\rangle_{magic} \otimes \left| 0 \right\rangle^{d-1}_{col} \otimes \left| \phi \right\rangle_{else} \nonumber\\
+ \beta \left| 1 \right\rangle_{magic} \otimes \left| 0 \right\rangle^{d-1}_{col} \otimes \left| \phi \right\rangle_{else}
\end{align}

\indent Similarly, by preparing all qubits in the same row as the physical magic state in the $+$ state, the expected value of $X_L$ aligns with that of the $X$ for the physical magic state:

\begin{align}
\left| \psi \right\rangle = \alpha \left| 0 \right\rangle_{magic} \otimes \left| 0 \right\rangle^{d-1}_{col} \otimes \left| + \right\rangle^{d-1}_{row} \otimes \left| \phi \right\rangle_{else} \nonumber\\
+ \beta \left| 1 \right\rangle_{magic} \otimes \left| 0 \right\rangle^{d-1}_{col} \otimes \left| + \right\rangle^{d-1}_{row} \otimes \left| \phi \right\rangle_{else}
\end{align}

After performing stabilizer measurements on this initialized state, the state is projected onto an eigenstate of the stabilizers. If all syndrome qubits are measured as 0, the logical quantum state $\left| M \right\rangle_L$ can be expressed as follows, where $S_i$ represents the $i$-th element of the set of stabilizer operators:

\begin{align}
\left| M \right\rangle_L &= \prod_{i} ( I + S_i ) \left| \psi \right\rangle \nonumber \\
&= \alpha \prod_{i} ( I+S_i ) \left| 0 \right\rangle_{magic} \otimes \left| 0 \right\rangle^{ d_1 -1}_{col} \otimes \left| + \right\rangle^{ d_1 -1}_{row} \otimes \left| \phi \right\rangle_{data}  \nonumber \\
&+ \beta \prod_{i} ( I+S_i ) \left| 1 \right\rangle_{magic} \otimes \left| 0 \right\rangle^{ d_1 -1}_{col} \otimes \left| + \right\rangle^{ d_1 -1}_{row} \otimes \left| \phi \right\rangle_{data} \nonumber \\
&= \alpha \left| 0 \right\rangle_L + \beta \left| 1 \right\rangle_L
\end{align}

As shown in the above process, there are various ways to initialize the remaining data qubits $\left| \phi \right\rangle_{data}$, excluding the physical qubit where the magic state is implemented, and the data qubits in the same row and column. However, various errors can occur during this process; therefore, it is essential to prepare the state of the data qubits in a way that facilitates error detection during stabilizer measurements to minimize errors.

\begin{figure}[t]
\centering
\includegraphics[width=1\linewidth]{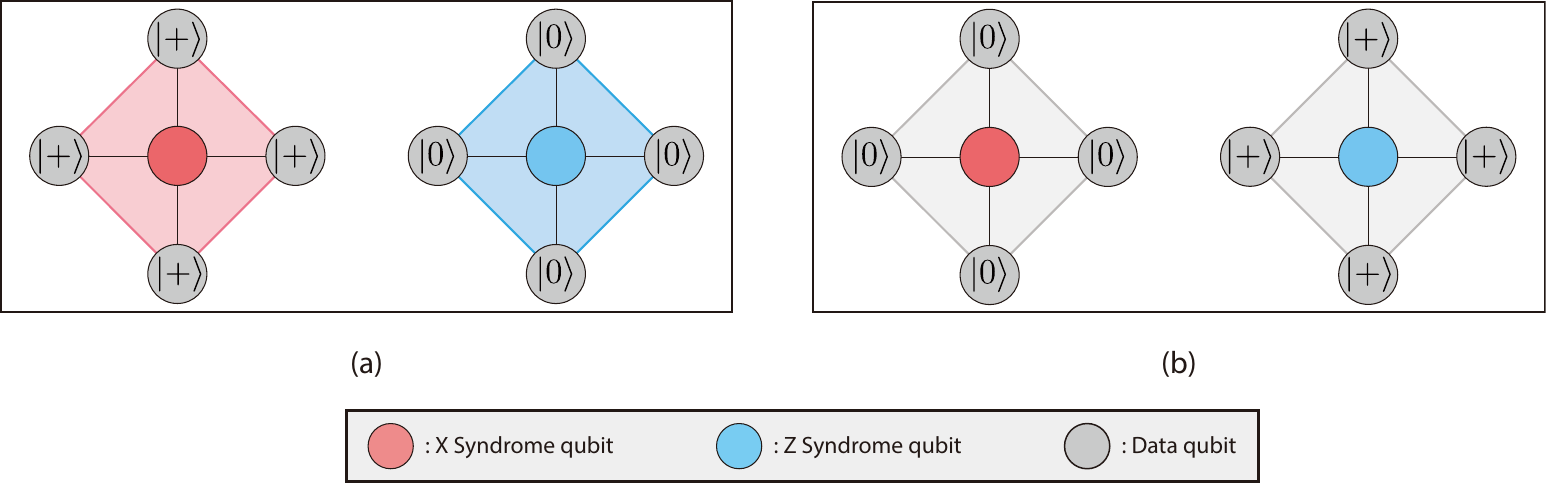} 
\caption{Types of stabilizer measurements. (a) Data qubits prepared as eigenstates of the stabilizer enable error detection without changing the state. (b) Data qubits are not initially eigenstates of the stabilizer. Quantum state collapse into an eigenstate during the measurement process. Errors occurring during this collapse are not detected by syndrome qubits, as the state changes during the measurement. }
\label{fig:stabilizer measurement}
\end{figure}

When performing stabilizer measurements in the quantum state, two distinct scenarios can arise (Figure \ref{fig:stabilizer measurement}). In the first scenario, the quantum state is already an eigenstate of the stabilizer. If no errors occur during the stabilizer measurement, the quantum state remains unchanged and the syndrome qubit is measured as 0. However, if an error occurs, the syndrome qubit is measured as 1, indicating the presence of an error. In the second scenario, the quantum state is not an eigenstate of the stabilizer. In this case, performing stabilizer measurements collapses the state into one of the eigenstates of the stabilizer operator. During this process, even if an error occurs, it cannot be detected because the state is altered as part of the measurement process.

To detect errors during magic state injection, it is crucial to perform stabilizer measurements that can detect errors on as many data qubits as possible. Specifically, these measurements should be performed on data qubits in the same row and column as the physical qubit where the magic state is prepared.  For effective error detection, the surrounding data qubits must be prepared in the same state. By initializing the remaining data qubits to either $|0\rangle$ or $|+\rangle$, depending on the states of the surrounding qubits, stabilizer measurements can be employed to detect errors. Since preparing qubits in the same state consistent with their surroundings enhances error detection, the data qubits should be divided into two regions: one prepared in the $|0\rangle$ and the other in the $|+\rangle$. Various methods can be used to determine the boundary between these two regions. The choice of initialization method during magic state injection significantly influences the logical error rate of the final logical magic state.

By following this procedure, a physical magic state can be encoded into a logical state, while simultaneously detecting errors through stabilizer measurements. However, errors that occur during this process cannot be corrected because the initially prepared state is not a logical state. Thus, errors are identified through post selection, a process that discards states with detected errors and retains those believed to be error-free. One limitation of this method is that, when using a larger number of physical qubits to create a logical state with a larger distance, the probability of obtaining an error-free state diminishes. To address this issue, a method is employed where a state injected into a smaller-distance code is extended to a larger-distance code. In this method, error correction becomes possible during the extension process, as a logical state with a smaller distance has already been obtained. Additional rounds of error correction must be performed to correct errors that occur during this process. This approach allows for the preparation of a larger-distance magic state while maintaining a reasonable probability of success. 

\subsection{Error Model}

\indent The types of errors that physically occur in real quantum computers vary widely, making it difficult to express them with precise mathematical formulations. To address this, error models that effectively simulate the behavior of actual quantum computers are needed. In this study, we conducted simulations using the depolarization error model and  Z-biased Pauli error model, as they closely represent the types of errors commonly occur in real quantum computers.

The depolarization error model is the most commonly used error model for evaluating the performance of error-correction techniques. In this model, errors occur with equal probability $p/3$ for Pauli $X$, $Y$, and $Z$. Additionally, we developed a Z-biased error model to account for the increased likelihood of Z-type errors in real quantum computers. In this model, the bias $\eta$ determines the probability of Z errors compared to X and Y errors. Thus the error probabilities for single-qubit and two-qubit gates should be adjusted based on the bias. For single-qubit gates, the error probabilities are defined as follows: 

\begin{align}
    P_X = P_Y = \frac{P_{\text{single}}}{2(\eta+1)}, \quad P_Z = \frac{2\eta P_{\text{single}}}{2(\eta+1)}
\end{align}
In this error model, when the bias $\eta = 0.5$, it corresponds to the depolarization error model where $P_X = P_Y = P_Z = P_{\text{single}}/3$. As $\eta$ increases to infinity, only Z errors occur with probability $p$. 

For the two-qubit gates, the depolarization error model results in each of the errors from the set $\{I, X, Y, Z\}^{\otimes 2}/\{I \otimes I\}$ a probability of $p/15$. In the biased case, the probabilities for errors involving Z type errors on one or both qubits ($ZZ$, $ZI$, and $IZ$) are increased by adjusting $\eta$ as follows:

\begin{align}
    P_{IX} = P_{IY} = P_{XI} = P_{XX} = P_{XY} = P_{XZ} = P_{YI} = P_{YX} = P_{YY} = P_{YZ} = P_{ZX} = P_{ZY} = \frac{P_{\text{double}}}{6(\eta+2)}
\end{align}

\begin{align}
    P_{ZZ} = P_{ZI} = P_{IZ} = \frac{P_{\text{double}}}{6(\eta+2)} + \frac{(2\eta-1)P_{\text{double}}}{6(\eta+2)} = \frac{\eta P_{\text{double}}}{3(\eta+2)}
\end{align}
When $\eta = 0.5$, this model becomes the depolarization error model where all errors occur with a probability of  $P_{\text{double}}/15$. As $\eta$ approaches infinity, the probabilities for $P_{ZZ}$, $P_{ZI}$, and $P_{IZ}$ errors converge to $P_{\text{double}}/3$, while the probabilities of all other errors become zero.

This error model represents the situation where increasing the bias decreases the occurrence of $X$and $Y$ type errors while increasing the rate of $Z$ type errors. In surface and XZZX codes, a $Y$ error is equivalent to simultaneous $X$ and $Z$ errors. Consequently, as the probability of $Y$-type errors decreases with increasing $\eta$, the likelihood of logical $X$-type errors is significantly reduced. However, the probability of logical $Z$ errors increase only to a lesser extent because the $Y$ error probability decreased. Therefore, owing to the reduction in $Y$ errors, the overall logical error rate tends to decrease as the bias increases.

The logical error rate was calculated to evaluate the error rate of the magic state. The logical qubit was measured in either the $X$ or $Z$ basis to determine the occurrence of logical error in the prepared state. By repeatedly performing the magic state injection and measuring in either the $X$ basis or the $Z$ basis, we obtained the logical $X$ and $Z$ error rates ($E_{X}$ and $E_{Z}$). The total logical error rate ($E_{total}$) was then calculated using the following formula, which accounts for cases where neither a logical $X$ nor $Z$ error occurred.

\begin{align}
    E_{\text{total}} = 1 - (1 - E_X) \times (1 - E_Z)
\end{align}

\section{Results}

To identify the optimal magic state injection strategy for the heavy-hexagon structure, we conducted experiments examining various characteristics. Specifically, we explored how the logical error rate changes with varying physical error rates in both the lattice and heavy-hexagon structures, comparing the surface code and two types of XZZX codes. In addition, we investigated the impact of varying error model biases, distances of the error correction codes, and initialization methods on the performance of each error correction code.

\subsection{Characteristic of Flag Qubit}

QEC codes generally use syndrome qubits to perform stabilizer measurements, where each syndrome qubit connects to a specific number of data qubits (referred to as the “weight”). However, hardware constraints often limit these connectivity requirements, making the direct application of error-correction codes impossible. To address this problem, flag qubits are employed. Flag qubits enable indirect connections between data and syndrome qubits, even under limited hardware connectivity. 

\begin{figure}[t]
\centering
\includegraphics[width=1\linewidth]{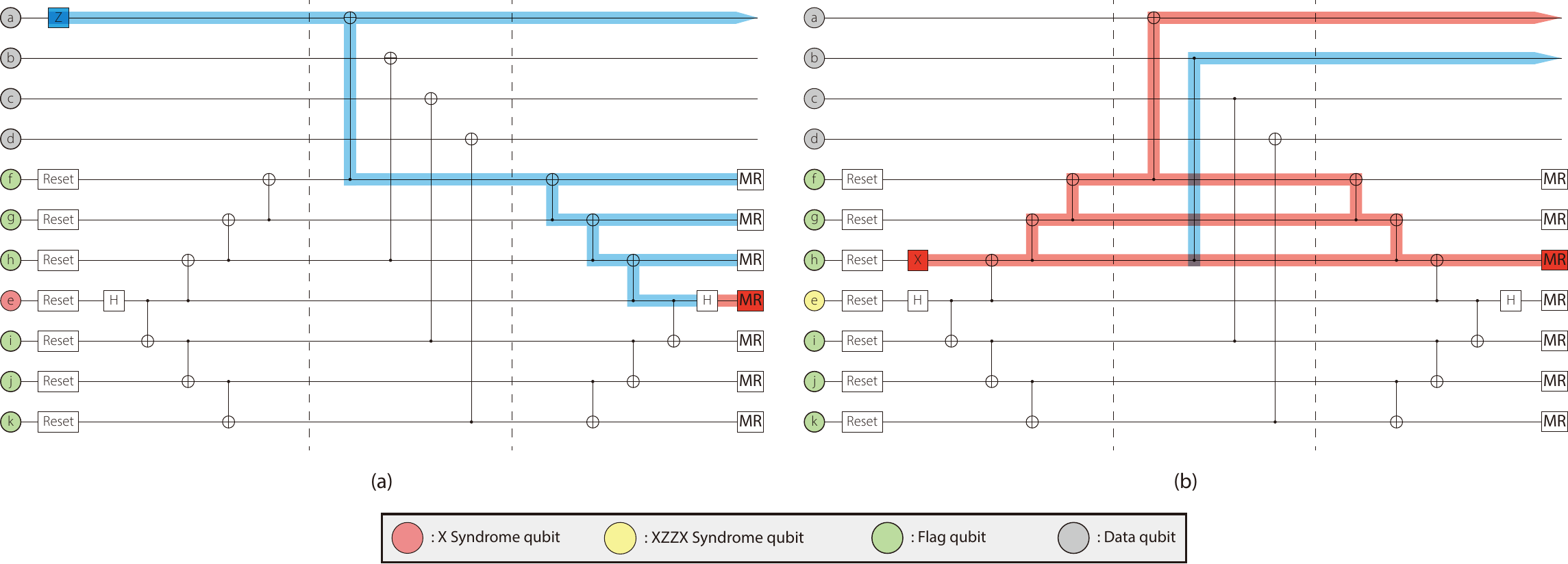} 
\caption{(a) Error propagation circuit for the X stabilizer on in the surface code implemented in the heavy-hexagon structure. Errors in data qubits propagate through flag qubits to syndrome qubits. Propagated Z-type errors do not affect the flag qubit measurement process, and are detected only from syndrome qubit measurements. (b) Error propagation circuit for the XZZX-type stabilizer in the XZZX code within the heavy-hexagon structure. An X error in a flag qubit propagates to data qubits as X or Z errors depending on the stabilizer via CNOT gates. These errors are undetectable by syndrome qubit measurements but can be identified through flag qubit measurements. Additional errors from flag qubits are a significant feature of the heavy-hexagon structure.}
\label{fig:error propagation}
\end{figure}

Figure \ref{fig:error propagation} shows a part of the stabilizer measurement circuits for the error correction codes implemented in the heavy-hexagon structure (Figure \ref{fig:surface code on HH with flag} and \ref{fig:XZZX code on HH with flag}). These figures illustrate that errors in data qubits propagate through the flag qubits and are subsequently measured by syndrome qubits. However, using flag qubits introduces additional physical qubits, which, in turn, can generate additional errors that must be detected and corrected.

Flag qubits are connected to nearby syndrome qubits, data qubits, and other flag qubits via CNOT gates. In this configuration, qubits closer to the syndrome qubit act as control qubits, while qubits closer to the data qubit serve as target qubits. The CNOT gates propagate X errors from the control qubit to the target qubit and Z errors from the target qubit to the control qubit. Therefore, if an X error occurs in a flag qubit, it propagates to the data qubits, whereas a Z error propagates to the syndrome qubits. Consequently, X errors in the flag qubits act as additional errors in the data qubits, whereas Z errors contribute to measurement errors in the syndrome qubits.

\begin{figure}[t]
\centering
\includegraphics[width=1\linewidth]{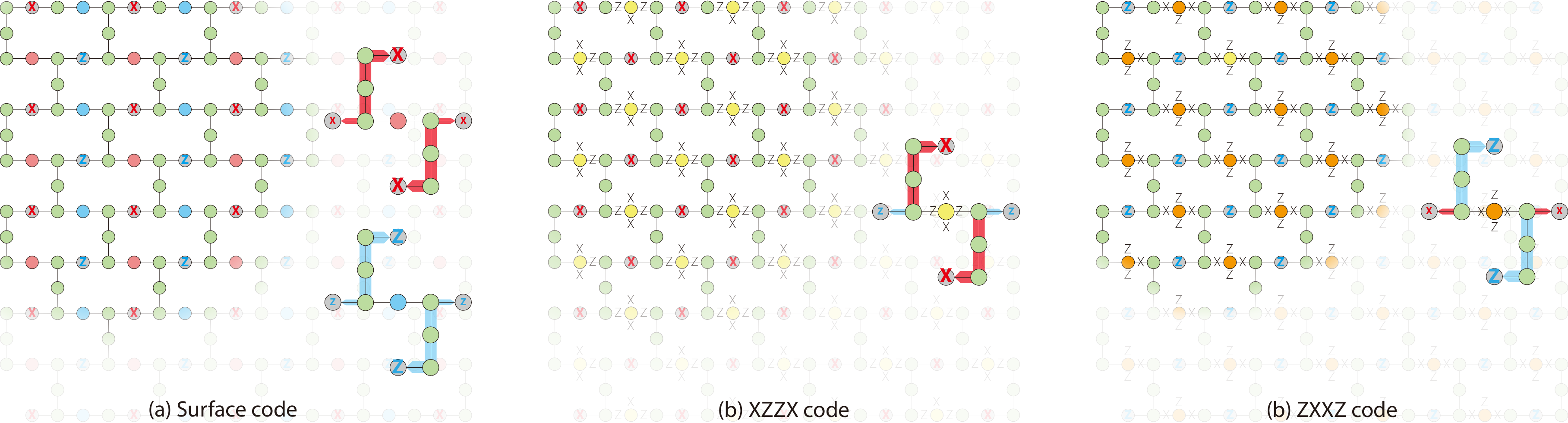} 
\caption{Illustration of additional errors caused by X errors in flag qubits. For the surface code, additional X or Z errors occur depending on the position of data qubits. For the XZZX code, different stabilizer types lead to more X errors in the XZZX type and more Z errors in the ZXXZ type across all data qubits. These additional errors introduce biases beyond those modeled by the error model.}
\label{fig:bias from flag qubit}
\end{figure}

The flag qubits are measured in each round to detect errors and reset their states. As the flag qubits are prepared in the $|0\rangle$ state after Z measurements, the Z errors propagated through the flag qubits do not affect the flag qubit measurement results. A Z error in the flag qubit, therefore, only affects the measurement outcome of the syndrome qubit and is considered a readout error. However, X errors in the flag qubits propagate to the data qubits, introducing errors dependent on the stabilizer operator. By measuring the flag qubits, it is possible to identify whether an X error has occurred. In addition, the CNOT gates involving flag qubits are symmetrically applied. This symmetry causes propagated errors to other flag qubits to cancel out, leaving only the error detected in the original flag qubit. This enables identification of which flag qubit experienced an X error based on the measurement results. Consequently, additional X errors from the flag qubits result in increased error rates in the data qubits. The effect of these errors varies depending on the stabilizer form and the position of the data qubits (see Figure \ref{fig:bias from flag qubit}). In the heavy-hexagon structure, the errors in each data qubit exhibit varying biases due to X errors in the flag qubits. Increasing the Z bias in the heavy-hexagon structure reduces the occurrence of X errors in the flag qubits, which, in turn, lowers the logical error rate. This bias-dependent behavior further decreases the logical error rate beyond the effects predicted by the error model. 

\subsection{Qubit Initialization}

The logical error rate of the resulting magic state depends on how each physical qubit is prepared in a physical state during magic state injection. We investigated the case where the physical magic state was prepared on the top-left qubit. In this configuration, the data qubits in the top row and leftmost column must be initialized to either the $|0\rangle$ or $|+\rangle$, depending on the logical operator determined by the given error correction code. The remaining data qubits should be prepared to minimize initialization errors while allowing effective error detection.

If the data qubits corresponding to the Pauli Z of a given stabilizer are initialized to the $|0\rangle$, and those corresponding to the Pauli X are initialized to the $|+\rangle$, errors arising during the initialization process can be detected. However, it is impossible to initialize data qubits in this manner for all stabilizers. To address this, syndrome qubits must be selected to detect errors effectively. Reducing logical errors during the magic state injection requires maximizing the number of syndrome qubits capable of detecting initialization errors. Achieving this necessitates an optimal division of regions where data qubits are prepared in either the $|0\rangle$ or $|+\rangle$. These regions can be classified in several ways.

\begin{figure}[!ht]
\centerline{\includegraphics[width=0.7\linewidth]{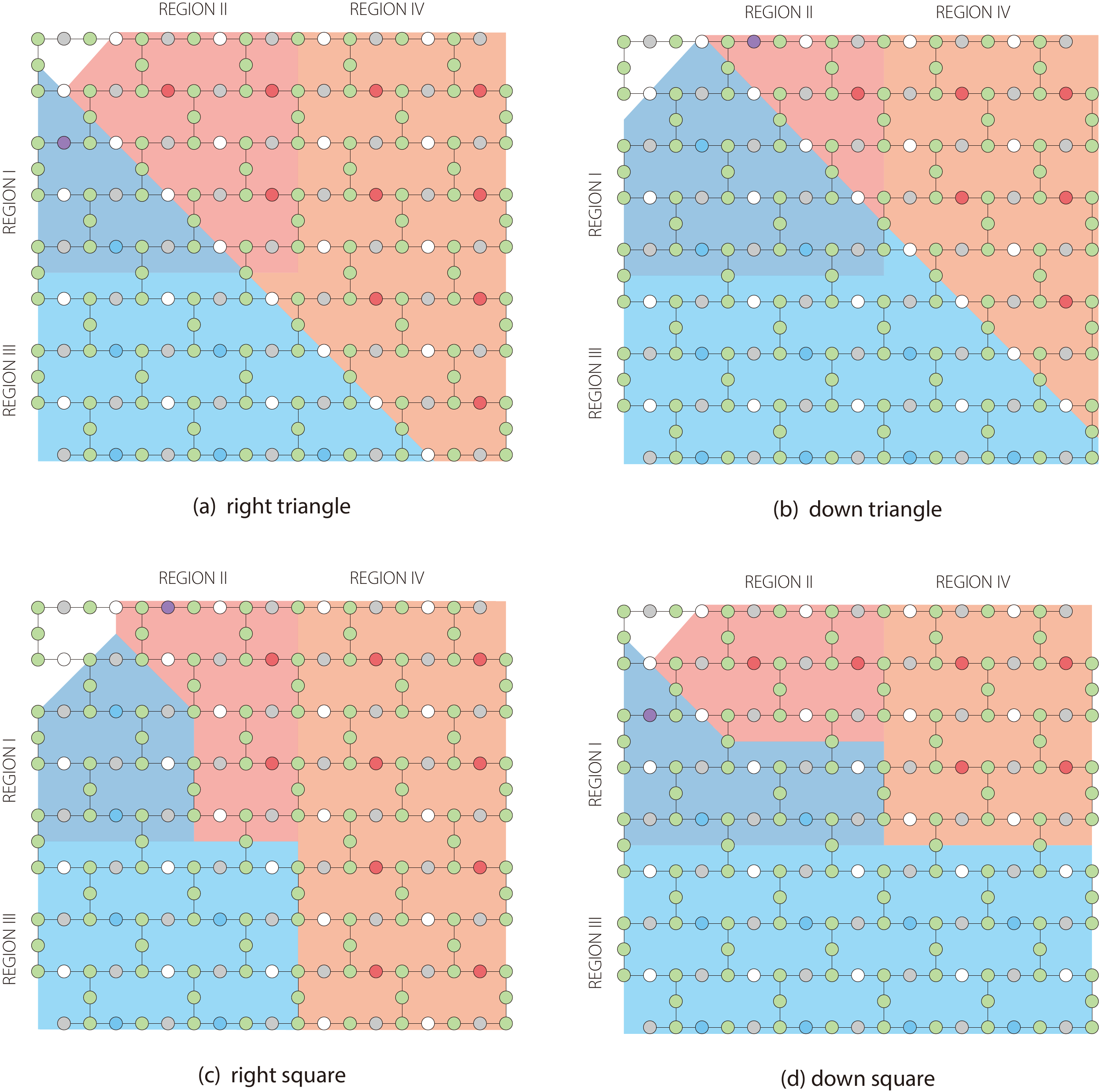}}
\caption{Four types of qubit initialization. (a) Right triangle method, (b) down triangle method, (c) right square method, (d) down square method. "Down" indicates the case where the number of data qubits in regions I and III (blue areas) is larger, and "right" indicates the case where the number of data qubits in regions II and IV (red areas) is larger. Data qubits in each area are initialized to the same state, either $|0\rangle$ or $|+\rangle$. The blind qubit is shown in purple, and the syndrome qubits capable of detecting initialization errors are depicted in red and blue. The down and right methods are symmetric with respect to the diagonal running from the top-left to the bottom-right. The square method divides initialization regions with a vertical or horizontal line, while the triangle method divides the regions using a diagonal line from the top-left to the bottom-right. The syndrome qubits that can detect errors are marked in deep red or blue. Regardless of the selected method, there is always one data qubit (the blind qubit, shown in purple) for which initialization errors cannot be detected. The position and initialization state of the blind qubit significantly influence the logical error rate. The blind qubit occupies the same position in the down square and right triangle methods. Similarly, it occupies the same position in the right square and down triangle methods, but its position differs between these two pairs. Consequently, these pairs of initialization methods exhibit similar logical error rate trends.}
\label{fig:qubit initialization}
\end{figure}

We conceptualized two main schemes for initializing the data qubits. The triangle scheme divides the initialization regions using a diagonal line from the top-left to the bottom-right, separating areas where data qubits are initialized to the $|0\rangle$ and $|+\rangle$. The square scheme divides the regions using a straight line, either from top to bottom or from left to right. In the square scheme, using a single straight line to divide the regions reduces the number of stabilizer measurements available for error detection. To address this issue, we adjusted the boundaries around the physical qubit where the magic state was injected. Each scheme is further divided into two methods depending on which of the two regions contains more data qubits: if the right region contains more data qubits, it is classified as "right," and if the lower region contains more data qubits, it is classified as "down." Thus, we performed simulations for the four initialization methods: right square, down square, right triangle, and down triangle (see Figure \ref{fig:qubit initialization}).

When the physical magic state is placed in the top-left qubit, there is always one data qubit for which errors cannot be detected. From here on, we will refer to this qubit as the 'blind qubit.' If errors propagate to this blind qubit during initialization and projection measurements, they can lead to logical errors. Therefore, it is crucial to carefully examine the errors that occur in the blind qubits. The errors affecting a blind qubit depend on its relative position and physical state. The relative position of the blind qubit changes when the "right" and "down" regions are swapped. In addition, the physical state of the blind qubit may vary depending on the error-correction code. We divided the XZZX code into two types: XZZX type and ZXXZ type. For a ZXXZ-type stabilizer, the logical operator differs, requiring different physical states for preparation. Consequently, for the ZXXZ type, the types of data qubits prepared in each region will change between $|0\rangle$ and $|+\rangle$. However, the XZZX and ZXXZ types are symmetric with respect to the diagonal running from top-left to bottom-right. Since the four initialization methods also have symmetric structures, if both XZZX-type and ZXXZ-type codes use symmetric initialization methods, the two will become equivalent. To ensure a fair comparison, we should compare the right-initialization of the XZZX type with the down-initialization of the ZXXZ type and the down-initialization of the XZZX type with the right-initialization of the ZXXZ-type.

Errors in the blind qubit can remain undetected and contribute to the logical error rates in two main scenarios: during the preparation of the physical state of the blind qubit and when errors propagate during the first stabilizer measurement process.  Errors arising from preparing the initialization of data qubits are basically single-qubit errors due to the application of Hadamard gates, which generally result in higher error rates when preparing the $|+\rangle$ compared to the $|0\rangle$. However, we assumed that two-qubit errors are 20 times more significant than single-qubit errors. This effect, therefore, is considerably less significant than the errors arising from the first stabilizer measurement, which involves a two-qubit gate. During the first projection measurement, undetected errors in the stabilizer measurements do not affect the logical error. However, errors in the stabilizer measurements that can detect errors may affect the logical error rate. In particular, errors in the syndrome or flag qubits can propagate to data qubits, potentially introducing additional errors beyond those originating from the initialization process. Therefore, the relative position of the syndrome qubits to the blind qubit during error-detecting stabilizer measurements can affect the logical error rate. If errors from the syndrome or flag qubits propagate to the blind qubit during stabilizer measurements, they will go undetected and directly increase the logical error rate. To better understand this effect, we need to examine how errors propagate to the blind qubits during initialization. If the blind qubit is located below the qubit where the magic state is prepared (e.g., in the right triangle or down square methods), the error is more likely to propagate through the syndrome qubits on the right. Conversely, if the blind qubit is positioned to the right of the magic state preparation qubit (e.g., in the down triangle or right square method), the error is more likely to propagate from the syndrome qubits located below. In the lattice structure, errors propagating from the blind qubit do not exhibit significant differences based on the direction because the horizontal and vertical directions of the stabilizers are equivalent. Therefore, the differences in single-qubit gate error rates due to the initialization state are more prominent. However, in a heavy-hexagon structure that uses more flag qubits in the vertical direction, a blind qubit located to the right (e.g., in the down triangle or right square) experiences greater error propagation, making the position of the blind qubit more influential.

\begin{figure}[!t]
\centering
\includegraphics[width=0.93\linewidth]{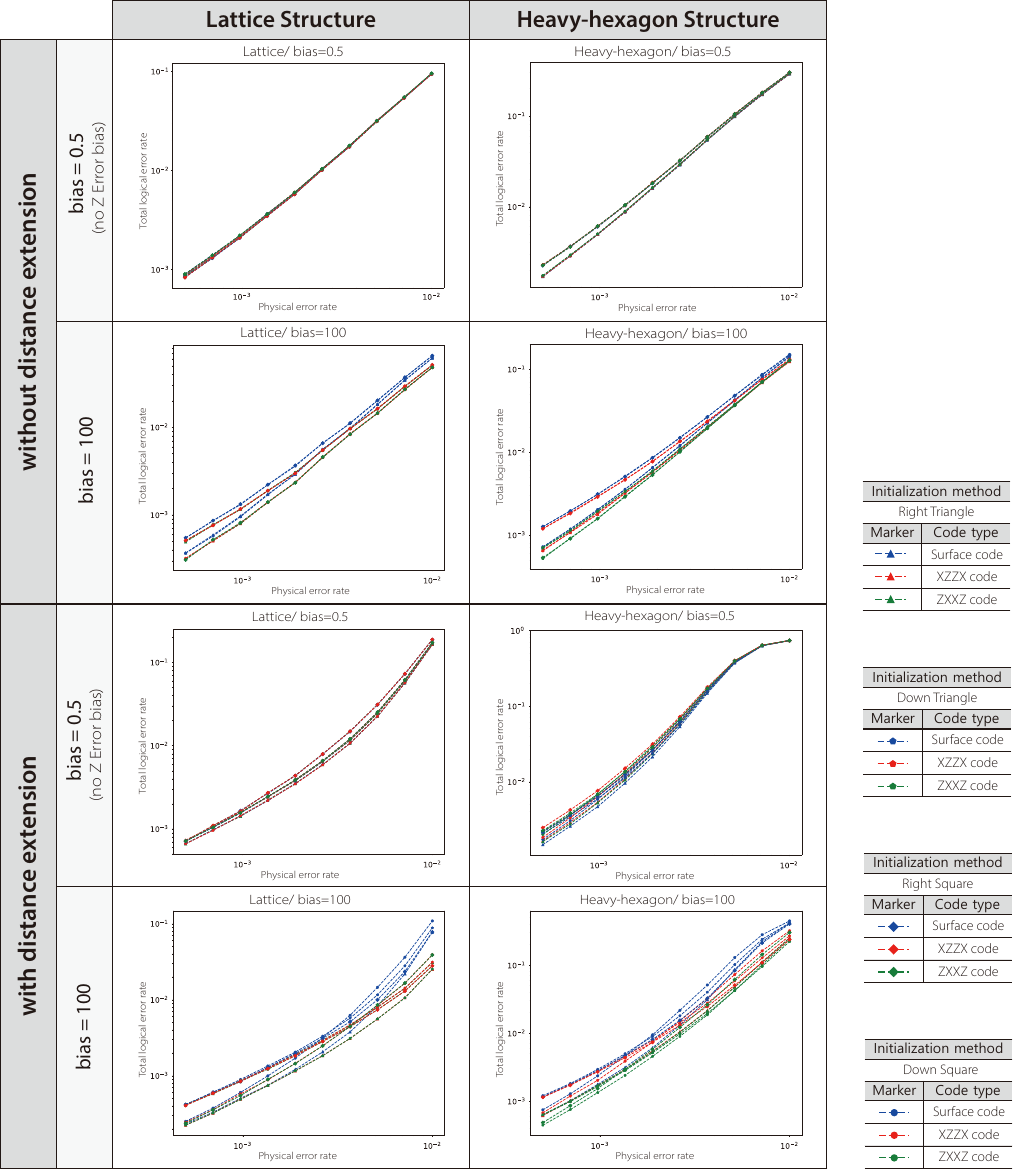}

\caption{Top: Comparison of logical error rates for heavy-hexagon and lattice structures with a bias of 0.5 and 100 without distance extension. Bottom: Comparison of logical error rates for heavy-hexagon and lattice structures with a bias of 0.5 and 100, with distance extended to 9. In the lattice structure, logical error rates are almost the same regardless of the initialization methods without bias or distance extension. In contrast, for the heavy-hexagon structure, the logical error rate varies based on the position of the qubit that cannot detect errors. As the bias increases, XZZX codes show lower logical error rates compared to the surface code. While the two XZZX code types show the same logical error rates for equivalent initialization in the lattice structure, the ZXXZ type shows lower error rates in the heavy-hexagon structure. Additionally, the triangle-based initialization methods tend to result in lower error rates overall when distance extension is applied.}
\label{fig:initialization graph}
\end{figure}

Figure \ref{fig:initialization graph} presents the results of our experiments using different initialization methods. We compared four initialization methods and three error-correction code types to determine the optimal combination that minimizes the error rate in a heavy-hexagon structure. To achieve this, we observed changes in the logical error rates during magic- state injection with varying bias and distance extension. Initially, we verified the logical errors in the lattice structure without applying bias or distance extension. All initialization methods showed similar error rates, with slight differences due to additional Hadamard gates used when the blind qubit was initialized to the $|+\rangle$. In the heavy-hexagon structure, we observed two distinct error rates, depending on the location of blind qubit relative to the magic state preparation qubit. Errors propagated through the syndrome qubits when the blind qubit was positioned to the right, leading to higher error rates than when it was positioned below the preparation qubit. As the bias increased, the XZZX code exhibited lower logical error rates due to the bias. In the lattice structure, the XZZX and ZXXZ codes with symmetric initializations exhibited similar error rates, with differences based on the initialization state of the blind qubit. If the blind qubit was initialized in the $|0\rangle$, an X error caused a logical error, while if it was initialized in the $|+\rangle$, a Z error resulted in a logical error. Therefore, as the Z bias increased, a lower error rate was observed when the blind qubit was initialized in the $|0\rangle$. This trend also maintained in the heavy-hexagon structure; however, as the bias increased, the ZXXZ type exhibited a lower error rate than the XZZX type (as explained in the following section).

When no distance extension was applied, the differences in logical error rates based on the initialization methods were minimal, due to the similar sizes of the regions. However, as the distance increased, the differences in the number of data qubits in each region became more significant, leading to larger variations based on the boundary configurations. At larger distances, the square method involves data qubits prepared in the same physical state across a larger area. This increases the likelihood of a chain of errors, leading to a higher logical error rate. Additionally, methods with larger regions of $|+\rangle$ qubits (e.g., the right method for surface code or XZZX type) required additional Hadamard gates, which introduced more errors. This effect became more pronounced with increasing distance, with the right square (or down square for the ZXXZ type) showing the highest error rates and the down triangle (or right triangle for the ZXXZ type) showing the lowest error rates in the lattice structure. In the heavy-hexagon structure, the influence of flag qubits led to different error rates for each initialization method and error-correction code combination. Due to the presence of flag qubits, the number of flag qubits that propagate errors to the blind qubits can vary based on the relative position of the blind qubit. This variation results in different logical error rates for each initialization method in the heavy-hexagon structure. Based on our results, with bias and distance extension, the most suitable error-correction code and initialization method for the heavy-hexagon structure was the ZXXZ type with the down-triangle method. 
   
\subsection{Effect of Bias}

Our previous experiments demonstrated that lattice and heavy-hexagon structures exhibit different responses to changes in bias. Generally, due to the characteristics of the error model, an increase in bias leads to a decrease in the logical error rate. In the heavy-hexagon structure, the reduction in X errors in the flag qubits results in a more significant decrease in the logical error rate as the bias increases. Furthermore, in the stabilizer patches of the physical qubits, the number of flag qubits required for connections varies depending on whether the data qubit is positioned vertically or horizontally relative to the syndrome qubit. Therefore, in the heavy-hexagon structure, the error rates depend on the relative positions of the data qubits. In the surface code, the influence of flag qubits leads to different types of errors, depending on the location of the data qubits. Each data qubit acquires additional X or Z errors based on the type of syndrome qubit located above or below it. For example, if the syndrome qubits in the vertical direction correspond to X stabilizers and those in the horizontal direction correspond to Z stabilizers, the data qubit is more likely to experience X errors. In the surface code, data qubits with higher X errors and those with higher Z errors are present in similar proportions, resulting in no additional bias across all data qubits. In contrast, the XZZX code uses different operators for the stabilizers in the vertical and horizontal directions, and only one type of stabilizer is used. Consequently, errors corresponding to the stabilizer's vertical operator are more frequently propagated to all data qubits. Thus, using the XZZX-type stabilizer introduces additional X errors in the data qubits, whereas using the ZXXZ stabilizer results in additional Z errors, creating an additional bias beyond that set in the error model. (See Figure \ref{fig:bias from flag qubit}) This effect of flag qubits leads to different performances for the XZZX and ZXXZ types in the heavy-hexagon structure. In particular, in the presence of Z bias, the ZXXZ type performs better because it reinforces the Z bias. Therefore, when using flag qubits, the presence of additional bias errors in the data qubits results in different characteristics compared with the lattice structure, leading to different outcomes for the magic state injection in the heavy-hexagon structure.

\begin{sidewaysfigure}
    \centering
    \includegraphics[width=1\linewidth]{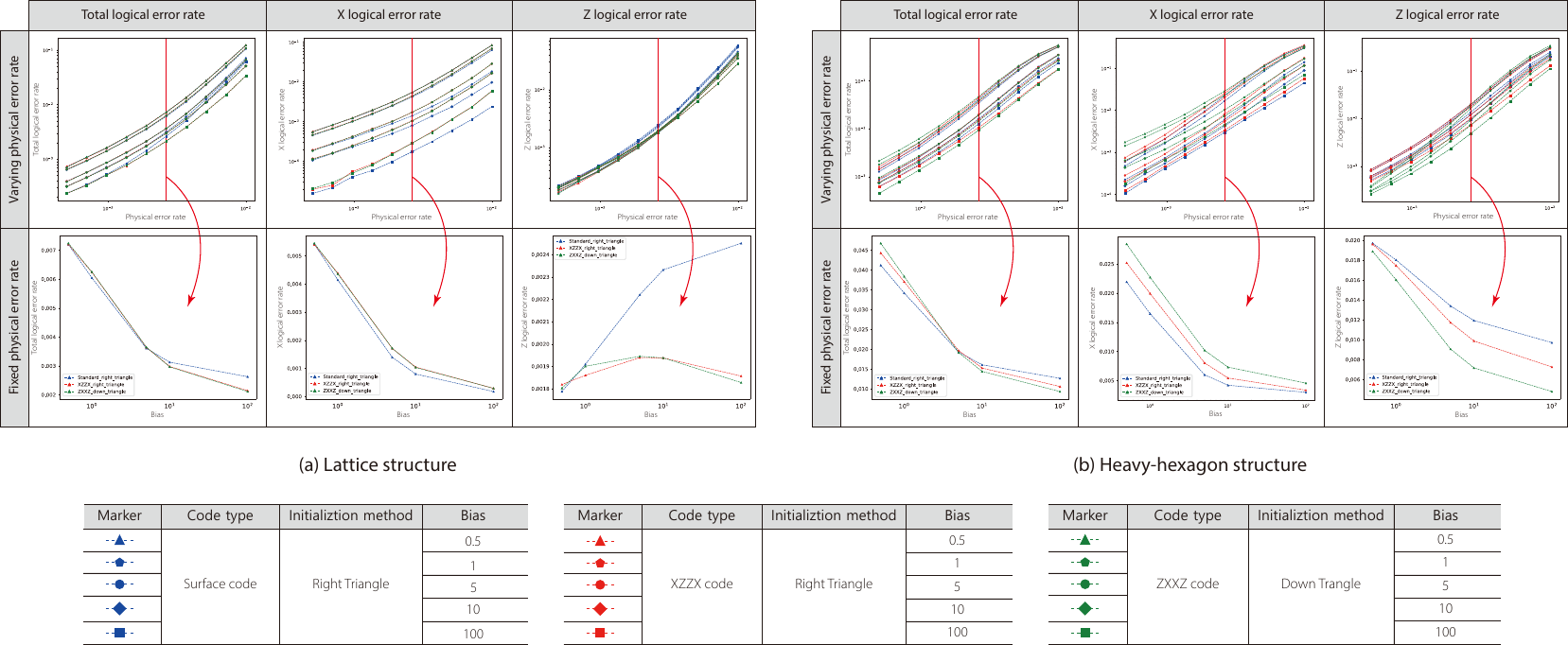}
    \caption{Graph showing how the logical error rate changes with varying bias. (a) Logical error rate with varying bias in a lattice structure. (b) Logical error rate with varying bias in a heavy-hexagon structure. The upper graph depicts changes in logical error rates as physical error rates vary for surface codes (XZZX and ZXXZ types) under different biases. The lower graph shows the changes in logical error rates as biases vary for each error-correction code at a specific physical error rate from the upper graph. Both heavy-hexagon and lattice structures show decreased total logical error rates and logical X error rates as bias increases. However, in the heavy-hexagon structure, the logical Z error decreases, while it increases in the lattice structure, likely due to the reduction of X errors in flag qubits as bias increases.}
    \label{fig:bias varying graph}
\end{sidewaysfigure}

We conducted magic state injection experiments using an error model with bias in both the lattice and heavy-hexagon structures for the XZZX and ZXXZ types, as well as the surface codes. As expected from the characteristics of the error model, both structures exhibited a decreasing trend in the logical error rate with increasing bias. We also confirmed that the XZZX and ZXXZ types exhibited lower logical error rates than the standard surface codes when the Z-bias was high. In the lattice structure, the error rates for the error-correction codes showed relatively small differences when there was no bias. ($\eta=0.5$). However, in the heavy-hexagon structure, even without bias, the presence of errors propagating from flag qubits caused the logical error rate to vary depending on the type of error-correction code. As the bias changed, the total logical error rate and the logical X and Z error rates changed differently in the lattice and heavy-hexagon structures. For the surface code in the lattice structure, increasing the Z bias decreases the logical X error rate while increasing the logical Z error rate. This outcome is consistent with the expected changes in physical qubit error rates. In contrast, for the XZZX and ZXXZ types in the lattice structure, the logical Z error increases more slowly or even decreases compared to the surface codes as the Z bias increases. This phenomenon occurs because the XZZX and ZXXZ types can detect errors more effectively using additional bias information. However, in the heavy-hexagon structure, an increase in the bias tends to decrease both the logical X and Z error rates across all error-correction codes, which is different from the lattice structure. This behavior results from the nature of errors in the flag qubits. In the heavy-hexagon structure, X errors in the flag qubits propagate to the data qubits, introducing additional errors. As the Z bias increases, X errors in the flag qubits decrease relatively, leading to a lower error rate in data qubits, and thus, a lower logical error rate. Furthermore, the difference between the XZZX and ZXXZ types becomes more pronounced with increasing bias due to differences in the types of errors propagated from flag qubits to data qubits. In the XZZX type, X errors are propagated more extensively through the three flag qubits, whereas in the ZXXZ type, Z errors are propagated more extensively. This results in each code experiencing a different effective bias, leading to different total error rates. Because the ZXXZ type introduces additional Z errors, it reinforces the Z bias, whereas the XZZX type introduces X errors, thereby reducing the Z bias. Consequently, the ZXXZ-type XZZX code exhibits the lowest logical error rate as the Z- bias increases.

\subsection{Error from Extension}

\indent The logical error rate after the magic state injection can be attributed to two main types of errors. The first type arises from errors during the initialization and projection measurements of the logical state. To mitigate these errors, we use a post selection process in which any state with detected errors is discarded. However, certain combinations of errors may go undetected, resulting in logical errors in the magic state. In addition, errors occurring in specific data qubits, such as those preparing the physical magic state or the blind qubit, may not be detected, contributing to the logical error rates from these qubits. The second type of error occurs during the process of increasing the distance and performing additional rounds, which can also lead to logical errors.  \\
\indent For the first type of error, the performance of the error-correction code was significantly influenced by the initialization method and the number of physical qubits used. In contrast, the second type of error can be corrected if the error-correction code is effective in detecting and correcting errors. Therefore, if the physical error rate is sufficiently low, the error-correction code can handle the secondary errors that arise during the distance extension process. However, this error reduction does not eliminate first-type errors from the initialization process. Therefore, reducing the physical error rate alone cannot infinitely reduce the errors that occur during the magic state injection process.

\begin{figure}[t]
\centering
\includegraphics[width=1\linewidth]{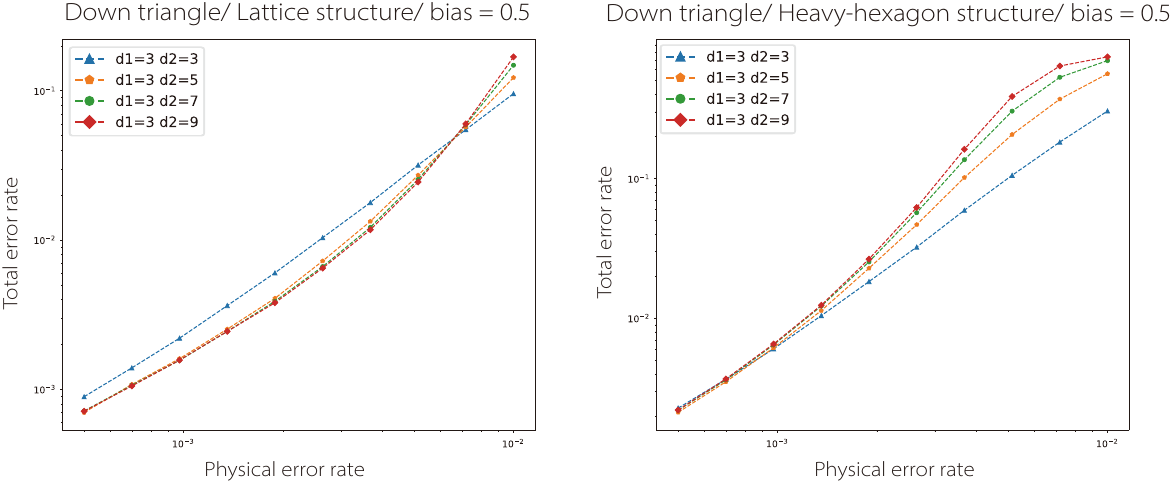} 
\caption{Logical error rates when extending a logical state prepared with a distance-3 error correction code to a distance-9 code. The figure shows results for the down- triangle initialization and ZXXZ error- correction code. In the heavy-hexagon structure, error rates according to different extensions become nearly identical only in the region of low physical error rates. However, in the lattice structure, logical error rates according to different extensions behave similarly below a certain threshold.}
\label{fig:Extenstion graph}
\end{figure}

We examined how the logical error rate of the magic state changed as distance increased. We conducted experiments to observe the logical error rate when extending from a distance-3 code ($d_1=3$)to a larger distance $d_2$. For comparison, we also conducted experiments with $d_1 = d_2$ (no distance extension). Each experiment involved injecting states into an error-correction code with distance $d_1$ and then extending the distance to $d_2$. We compared the logical error rates of the heavy-hexagon and lattice structures. In the case of the heavy-hexagon structure, for low physical error rates, the logical error rates showed similar behavior regardless of the distance. Above the low physical error regime, the logical error rate increased as the extension size increased. However, in the lattice structure, below a certain threshold, increasing the extension size reduced the logical error rates, whereas above this threshold, larger distances caused a more rapid decline in error rate \\
\indent This suggests that, with sufficiently good hardware, most errors occurring during distance extension can be corrected, and logical magic states with larger distances can achieve a similar logical error rate. However, some errors remain inevitable during the initial process, indicating that a distillation process is necessary to further reduce these errors.

\section{Conclusion}

Through these experiments, we demonstrated how magic state injection can be effectively implemented in the heavy-hexagon structure. Additionally, we confirmed that the presence of flag qubits introduces several phenomena not observed in the lattice structure during the magic state injection process. Specifically, the flag qubits in the heavy-hexagon structure contribute additional bias errors in the data qubits, which vary depending on the stabilizers. This observation indicates that the two types of stabilizers in the XZZX code exhibit different performances. Additionally, the initialization method--particularly the selection of the blind qubit position and the direction of the boundary lines--affects the logical error rate during the magic state injection process. Considering these factors, we conclude that the most suitable magic state injection method for the heavy-hexagon structure is to use the ZXXZ-type XZZX code combined with the down-triangle initialization method. This study emphasizes the critical considerations for identifying appropriate magic state injection methods for hardware with flag qubits. These findings are expected to facilitate the implementation of FTQC, reducing error rates in non-Clifford gates on such hardware. 

\section*{Data availability}
All data generated or used during the study appear in the submitted article.

\section*{Acknowledgement}
We thank Younghun Kim for his contribution in early stage of this work. This work is supported by the Basic Science Research Program through the National Research Foundation of Korea (NRF) funded by the Ministry of Education, Science and Technology (NRF2022R1F1A1064459) and Creation of the Quantum Information Science R$\&$D Ecosystem (Grant No. 2022M3H3A106307411) through the National Research Foundation of Korea (NRF) funded by the Korean government (Ministry of Science and ICT).





\section*{Appendix A: Magic State Injection Procedure}
Magic state injection is the process of encoding a physical magic state prepared in physical qubits into a logical magic state in logical qubits. The magic state injection process consists of two main stages. In Stage 1, a magic state is injected into a logical qubit of distance  $d_1$. Errors that occur during this process are detected, and post-selection is applied to obtain a logical magic state with no detected errors. In Stage 2, an error correction code with a larger distance $d_2$, is constructed, incorporating the physical qubits from Stage 1. Additional rounds of error detection and correction are performed on the larger-distance logical qubit. Following this, the data qubits are measured to evaluate the performance of the magic-state injection process.\\
\indent In this experiment, we set $d_1$ to 3 and varied $d_2$ as 3, 5, 7, 9 with biases $\eta$ of 0.5, 1, 5, 10, and 100. The physical error rate variable used refers to both the two-qubit error rate and the readout error rate. The probability of two-qubit gate errors varied from 0.05\% to 1\%. For simulation purposes, the probability of single-qubit gate errors was assumed to be 1/20 of the two-qubit gate error probability. Simulations were conducted using Stim code \cite{Gidney} and decoding was performed using the Pymatching algorithm. The following types of gates were used: CZ, CNOT, Hadamard, and computational measurements. \cite{Higgott} Sampling was done $1 \times 10^7$ times.

\subsection*{Stage I}

\begin{itemize}
	\item
	The first step in Stage I is determining the positions of the physical qubits that prepare the magic state. The positioning affects the subsequent configuration of the region. In the experiment, the data qubit was placed in the upper-left corner.
    \item 
    Next, we divide the region to initialize the data qubits. This region is generally divided into four parts. Regions I and II are distinguished based on whether the data qubit in the distance $d_1$ region is initialized to either the  $|0\rangle$ or $|+\rangle$. Regions III and IV are distinguished based on whether the data qubit in the distance $d_2$ region, which is not included in Regions I and II, is initialized to either the $|0\rangle$ or $|+\rangle$. We configured the regions such that Regions I and III were initialized to the same quantum state, and Regions II and IV were initialized to the same quantum state. In the special case where $d_1 = d_2$, Regions III and IV were not defined. The initialization of data qubits to $|0\rangle$ or $|+\rangle$ depends on how the logical X and Z operators are defined for the error correction code. For example, in the surface code or XZZX-type XZZX code, the data qubits in Regions I and III are initialized to $\left| + \right\rangle$, whereas those in Regions II and IV are initialized to $\left| 0 \right\rangle$. Conversely, for the ZXXZ-type XZZX code, the data qubits in Regions I and III were initialized to $\left| 0 \right\rangle$, and those in Regions II and IV were initialized to $\left| + \right\rangle$.
    \item
    Stabilizer measurements are then performed to collapse the prepared state into an eigenstate. Stabilizer measurements involving data qubits already prepared in the eigenstate of the stabilizer operator can detect errors that occur during the initialization of the data qubits. Using the results from these specific stabilizer measurements, we can detect errors occurring during the initialization process. However, due to issues such as readout errors that may arise during measurement, a single round of stabilizer measurement is not sufficient to confirm whether the projected state is logical state.
	\item
	Therefore, we performed two rounds of stabilizer measurements to verify whether errors occurred during the encoding process of the physical state into a logical state. If any error was detected in either round, the state was discarded via a post-selection process. Errors occurring in the flag qubits could be verified by measuring them and comparing their parities. After each stabilizer measurement step, the syndrome and flag qubits were reinitialized for the next round. This method allowed us to probabilistically obtain the $d_1$ distance logical state in Stage I.

\end{itemize}

\subsection*{Stage II}

\begin{itemize}
	\item
	The first step in Stage II is to initialize the data qubits in regions III and IV to achieve a distance $d_2$ magic state. This initialization is similar to the process used for Regions I and II , ensuring that Region III is initialized to the same state as Region I and Region IV to the same state as Region II.
	\item
	Stabilizer measurements are then performed across the entire $d_2$ distance region. Since the state in the $d_1$ distance region is already prepared as an eigenstate of the stabilizers, errors can be detected through the measurement results of the syndrome qubits. The stabilizer measurements in Regions III and IV project the prepared data qubit states onto the $d_2$ error correction code. In the $d_1$ distance region, the state of the data qubit is already an eigenstate of the stabilizer. This enables the identification of errors, including their locations and types, facilitating effective error detection and correction.
    \item
    After the projection measurements, we obtain the $d_2$ distance logical magic state. However, additional rounds are required to detect and correct errors that may have occurred during the measurement process. As more physical qubits are used and the distance increases, more rounds are required to check for errors in these qubits. We conduct $d_2$ rounds of stabilizer measurements for each logical qubit to detect errors that increase with the distance. In the case where $d_1 = d_2$, we do not perform additional qubit initialization, we instead add $d_1$ rounds of measurements and check the error rate for a fair comparison.
    \item 
    To determine the error rate of the $d_2$ distance logical magic state, the obtained state is measured. Possible logical errors include logical X, Y, and Z errors. Since a logical Y error can be considered as the simultaneous occurrence of logical X and Z errors, our analysis focused on detecting logical X and Z errors. To detect logical X and Z errors, all data qubits are measured on either the X or Z basis. Measuring the data qubits collapses the logical state to either the $|+\rangle$,$|-\rangle$ or $|0\rangle$,$|1\rangle$ depending on the measurement basis used. Using the measurement results, we employed the MWPM algorithm \cite{Edmonds,Kolmogorov,Fowler2,Fowler3} to identify the errors that occurred during the magic state injection process and calculated the appropriate correction operator. We analyzed the parity of the measurement results for the data qubits corresponding to the logical operators after applying the correction operator. If the parity of the logical X operator is 0 for measurements on the X basis, the logical state is considered to have collapsed to the $|+\rangle$. If the parity of the logical Z operator is 1 for measurements on a Z basis, then the logical state is considered to have collapsed to $|1\rangle$. By repeating this measurement process multiple times, we compute the expectation values for Pauli X and Z. Comparing these expectation values with those for the desired state allows us to calculate the probabilities of the logical X and Z errors. The total logical error rate is then calculated as follows:
    \begin{align}
    E_{total} = 1 - (1-E_{X}) \times (1-E_{Z})
    \end{align}
    
\end{itemize}
\indent Through this process, we can obtain an $\left| M \right\rangle_L$ of distance $d_2$ and implement non-Clifford logical operators in the heavy-hexagon structure.

\end{document}